\documentclass{article}
\usepackage{graphicx} 
\usepackage{amsmath}
\usepackage{amsfonts}
\usepackage{bm}
\usepackage{ragged2e}
\usepackage{float}
\usepackage{subcaption}
\usepackage{xcolor}

\usepackage [left = 2.5 cm, right = 2.5 cm, top = 2.5 cm, bottom = 2.5 cm, bindingoffset = 0 cm]{geometry}

\emergencystretch=\maxdimen
\title{Three-uniform edge-ordered hypergraph quantum states: entanglement and its relation to hypergraph properties}
\author{N. A. Susulovska, Kh. P. Gnatenko}
\date{}

\begin{document}

\maketitle

\begin{center}
	\emph{Ivan Franko National University of Lviv,\\
	Professor Ivan Vakarchuk Department for Theoretical Physics,\\
	12 Drahomanov St., Lviv, 79005, Ukraine}
\end{center}

\section*{Abstract}
The encoding of 3-uniform edge-ordered weighted hypergraphs into multiqubit quantum states is proposed. We derive an analytical expression for the geometric measure of entanglement of hypergraph quantum states and establish its relation to hypergraph properties. The results establish a direct connection between the local structure of hypergraphs, hyperedge weights, and the entanglement of the corresponding quantum states. For equally weighted hypergraphs, it is shown that the entanglement depends explicitly on the vertex degree. The dependence of entanglement on hypergraph parameters is studied analytically and using quantum computing. Hypergraph states associated with chain, star, regular lattice, and binary-tree hypergraphs are examined, and their entanglement is quantified using quantum computing. These results open up the possibility of studying hypergraph properties with quantum programming.

\section{Introduction}

Quantum graph states constitute an important class of entangled multiqubit quantum states that can be represented by graphs. These states have been extensively studied in recent years (see, for instance, \cite{Markham, Vesperini1, Vesperini2, Wang, Schlingemann, Mazurek, Mooney, Shettell, Gnatenko24, Gnatenko26, Susulovska} and references therein). The mapping between a quantum state and a graph structure is typically established by associating each vertex of the graph with a qubit and each edge between two vertices with the action of a two-qubit gate. Owing to their entanglement properties and convenient graph-based representation, graph states have found applications in a wide range of quantum information and quantum computing tasks, including quantum error correction \cite{Schlingemann, Mazurek}, quantum cryptography \cite{Markham,Qian}, quantum machine learning \cite{Gao,Zoufal}, and others.

A notion of hypergraphs emerges as a generalization of graphs, where
edges, now referred to as hyperedges, can join an arbitrary number of vertices. Formally, a hypergraph $H(E, V)$ is defined as a pair of a vertex set $V$ and a hyperedge set $E$. A hypergraph $H_k(V,E)$ is called $k$-uniform if all of its hyperedges contain exactly $k$ vertices. In these terms, graphs constitute a subset corresponding to 2-uniform hypergraphs $H_2(V, E)$.

In the same sense, hypergraph quantum states \cite{Huang, Rossi, Poderini, Guhne, Qu, Gachechiladze} are a generalization of quantum graph states. This generalization is achieved by allowing not only pairwise interactions between neighboring qubits but also entanglement between arbitrary subsets of qubits through hyperedges. This broader structure provides a more general class of entangled multiqubit quantum states.
The authors of \cite{Huang} experimentally prepared four-qubit hypergraph states on a fully reprogrammable silicon-photonic quantum chip and studied their entanglement properties. In \cite{Rossi}, a class of multiqubit quantum states associated with mathematical hypergraphs and generated using multiqubit controlled-Z gates was introduced. A one-to-one correspondence with quantum states used in algorithms such as Deutsch–Jozsa and Grover's algorithms was established. In \cite{Guhne}, the entanglement and nonclassical properties of hypergraph states were studied, and the local unitary equivalence of hypergraph states involving up to four qubits was investigated. The authors of \cite{Gachechiladze} showed that the correlations present in hypergraph states provide a basis for constructing different proofs of quantum nonlocality. A review of quantum hypergraph states was presented in \cite{Poderini}.

In our recent papers (see, for example, \cite{Gnatenko24, Gnatenko26, Gnatenko25, Susulovska} and references therein), we have studied quantum graph states and established relationships between the properties of these states and the structural properties of the corresponding graphs. As a result, methods for investigating graph properties using quantum computing have been proposed.

In the present paper, we introduce quantum states associated with 3-uniform edge-ordered weighted hypergraphs. The entanglement of these states is investigated both analytically and using quantum programming. Entanglement is one of the most important properties of quantum states that constitutes a resource for quantum computing \cite{Feynman, Horodecki, Horodecki1, Ekert, Lloyd}. 
We consider the geometric measure of entanglement, defined as
\begin{equation}
E(|\psi\rangle) = \min_{{|\psi_s\rangle}}\left(1 - |\langle \psi|\psi_s\rangle|^2\right),
\label{eq}
\end{equation}
where $|\psi\rangle$ is the entangled target state of interest, and
$d(|\psi\rangle, |\psi_s\rangle) = \sqrt{1-|\langle \psi|\psi_s\rangle|^2}$
is the Fubini--Study distance between $|\psi\rangle$ and a separable state $|\psi_s\rangle$. The minimization in (\ref{eq}) is performed over all possible separable states \cite{Shimony}.
A relationship between the entanglement of hypergraph quantum states and the properties of the underlying hypergraphs is established. These findings demonstrate the potential of quantum computing for investigating the structural properties of hypergraphs.

The paper is organized as follows. In Section \ref{s1}, the encoding of 3-uniform edge-ordered weighted hypergraphs into multiqubit quantum states is presented. The entanglement of quantum states representing edge-ordered weighted hypergraphs is studied analytically in Section \ref{s2}. An investigation of the dependence of entanglement on the parameters of hypergraph quantum states using quantum computing is presented in Section \ref{s3}. Section \ref{s4} is devoted to the conclusions.

\section{Three-uniform Edge-ordered Hypergraph States Generated by RZZY Operators}\label{s1}

Bulding on the established procedure for graph state construction, one can define a bijection between a set of hypergraphs containing hyperedges up to size $n$ and a class of multiqubit entangled states of the following structure

\begin{equation}
    |\psi_{HG}\rangle = \prod_{k=2}^n \prod_{(j_1,..,j_k)} U_{j_{1}...j_{k}} |\psi_{init}\rangle.
    \label{eq:general_hypergraph_state}
\end{equation}

\noindent According to this mapping, vertices of a hypergraph represent qubits prepared in the initial separable state $|\psi_{init}\rangle$, and hyperedges $(j_1,..,j_k) \in E$ of size $k$ indicate the presence of interaction within corresponding $k$-qubit clusters. This interaction is generated by the action of multiqubit entangling unitaries $U_{j_1...j_k}$ on the states of qubits with respective indices $j_1,...,j_k$. The product over $k$ in (\ref{eq:general_hypergraph_state}) accounts for hyperedges of different sizes and, hence, is discarded for the definition of $k$-uniform hypergraph states. 

Traditionally, hypergraph states obtained using multiqubit controlled-Z operators $C^kZ_{j_1...j_k}$ have been investigated. In this formalism, a system of $|V|$ qubits ($|V|$ being the cardinality of the vertex set of the associated hypergraph) is initialized in the uniform superposition state $|+\rangle^{\otimes |V|}$, where $|+\rangle = \left(|0\rangle + |1\rangle\right)/\sqrt{2}$ is the eigenstate of Pauli-$X$ operator. Subsequently, operators $C^kZ_{j_1...j_k}$ are applied to clusters of $k$ qubits, indexed $j_1,...,j_k$, where each cluster corresponds to a subset of adjacent vertices connected by a hyperedge of size $k$. Note that the action of $C^kZ_{j_1...j_k}$ on the computational basis results in the phase flip of state $|11...1\rangle_{j_1j_2...j_k}$, $C^kZ_{j_1...j_k} |11...1\rangle_{j_1j_2...j_k} = -|11...1\rangle_{j_1j_2...j_k}$, whereas all the other basis states remain unchanged. It was shown that the class of hypergraph states generated according to this rule is equivalent to the class of REW (real equally weighted) states, which are widely utilized in quantum computing as a resource for various quantum algorithms. Note that here the choice of interaction type introduced through $C^kZ_{j_1..j_k}$ unitaries is primarily dictated by the required uniqueness of hypergraph representation of an arbitrary state from class (\ref{eq:general_hypergraph_state}). To demonstrate that this condition is satisfied, a procedure for one-to-one hypergraph reconstruction of an arbitrary REW-equivalent hypergraph state has been proposed.

In the framework of this research, we construct multiqubit states associated with 3-uniform hypergraphs $H_3(V, E)$ with the help of parameterized operators 

\begin{equation}
    U_{ijk}(\theta_{ijk}) = RZZY_{ijk}(2\theta_{ijk}) = \text{e}^{-\text{i}\theta_{ijk}\sigma^z_i\sigma^z_j\sigma^y_k},
    \label{eq:unitary}
\end{equation}
where $\sigma^\alpha_l$, $\alpha \in \{x,y,z\}$, $l \in \{i,j,k\}$, is an appropriate Pauli operator acting on the qubit with index $l$, $\theta_{ijk} \in [0,2\pi)$, $(i,j,k)\in E$. Unitary (\ref{eq:unitary}) introduces a 3-qubit interaction in the system and forces a conditional rotation of the state of qubit $k$ around the Y-axis, where the direction of this rotation is determined by the states of qubits $i,j$

\begin{equation}
   RZZY_{ijk}(2\theta_{ijk})|x_ix_jx_k\rangle = |x_ix_j\rangle \otimes RY_k((-1)^{x_i+x_j}2\theta_{ijk})|x_k\rangle.
   \label{eq:rzzy_identity}
\end{equation}
Here $x_l \in \{0,1\}$, $l\in\{i,j,k\}$, are binary variables representing states of the respective qubits. 

Hypergraph states of interest are defined in the form

\begin{equation}
    |\psi_{HG}\rangle = \prod_{(i,j,k) \in E} \text{e}^{-\text{i}\theta_{ijk}\sigma^z_i\sigma^z_j\sigma^y_k} |+\rangle^{\otimes n},
    \label{eq:rzzy_hypergraph_state}
\end{equation}
where the number of qubits $n$ is equal to the cardinality of the vertex set $|V|$, $E$ is the set of hyperedges. Note, that for certain permutations $(i, j, k)$, operators (\ref{eq:unitary}) corresponding to adjacent hyperedges do not commute. Therefore, introduced class of multiqubit states facilitates the encoding of hypergraphs with edge ordering. From this perspective, our selection of the interaction type ensures one-to-one mapping between states (\ref{eq:rzzy_hypergraph_state}) and 3-uniform edge-ordered hypergraphs. Moreover, parameterization of the entangling unitaries allows us to encompass weighted hypergraphs, where parameters $\theta_{ijk}$ are assigned as weights to corresponding hyperedges. 

\begin{figure}[H]
  \centering
   \includegraphics[scale=0.45]{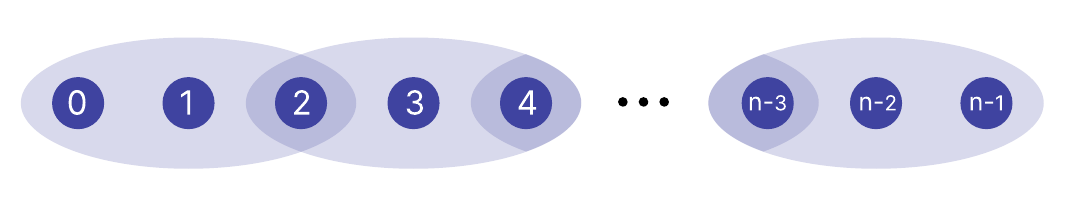}
    \caption{A 3-uniform chain hypergraph with single-vertex overlap between adjacent hyperedges representing multiqubit states of structure (\ref{eq:chain_state_general}), $\alpha=1$.} \label{fig:chain_single_hypergraphs}
\end{figure}

\begin{figure}[H]
  \centering
   \includegraphics[scale=0.45]{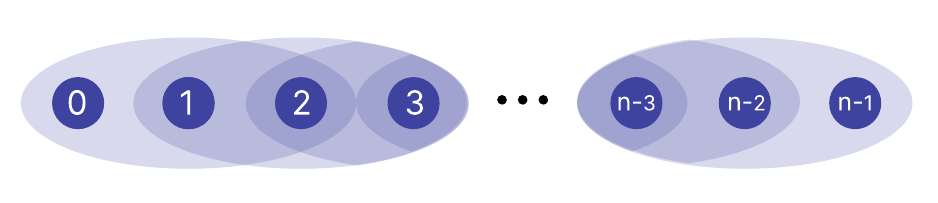}
    \caption{A 3-uniform chain hypergraph with double-vertex overlap between adjacent hyperedges representing multiqubit states of structure (\ref{eq:chain_state_general}), $\alpha=2$.} \label{fig:chain_double_hypergraphs}
\end{figure}

Within the class of hypergraph states under consideration, we focus on the subclasses corresponding to the hypergraphs with peculiar connectivity patterns. Namely, multiqubit states associated with chain- and star-like hypergraphs as well as more complex structures corresponding to regular lattices and binary trees are investigated. To begin with, let us consider 3-uniform chain hypergraphs with different connectivity. In particular, we examine cases where each pair of adjacent hyperedges in the chain intersects over a single vertex or two vertices (see Figures \ref{fig:chain_single_hypergraphs}, \ref{fig:chain_double_hypergraphs}, respectively). Corresponding quantum hypergraph states take the following form

\begin{align}\
    |\psi_{HG}^{chain,(alpha)}\rangle &= 
    \prod_{k=0}^{m-1}\text{e}^{-\text{i} \theta_\mathbf{{k}} \sigma^z_{(3-\alpha) k} \sigma^z_{(3-\alpha)k+1} \sigma^y_{(3-\alpha)k+2}} |+\rangle^{\otimes n},
    \label{eq:chain_state_general}
\end{align}
where $\alpha \in \{1, 2\}$ is introduced to distinguish between single- and double-vertex intersection patterns in the chain, index $\mathbf{k} = ((3-\alpha)k, (3-\alpha)k+1, (3-\alpha)k+2)$, $k \in \{0, ..., m-1\}$, runs over the set of hyperedges $E$ of cardinality $m$, parameters $\theta_\mathbf{k}$ determine the weights of corresponding hyperedges, $n=(3-\alpha)m + \alpha$ is the number of vertices in the hypergraph representing qubits of the multipartite quantum system. Note that in Figures \ref{fig:chain_single_hypergraphs}, \ref{fig:chain_double_hypergraphs} hyperedges of the chain hypergraphs are ordered from left to right, resulting in a unique correspondence to states (\ref{eq:chain_state_general}).

\begin{figure}[H]
  \centering
   \includegraphics[scale=0.45]{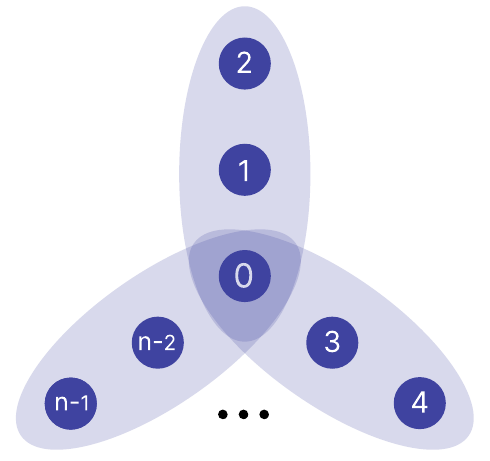}
    \caption{A 3-uniform star hypergraph with single-vertex overlap between adjacent hyperedges representing multiqubit states of structure (\ref{eq:star_single_state}).} \label{fig:star_single_hypergraphs}
\end{figure}

\begin{figure}[H]
  \centering
   \includegraphics[scale=0.3]{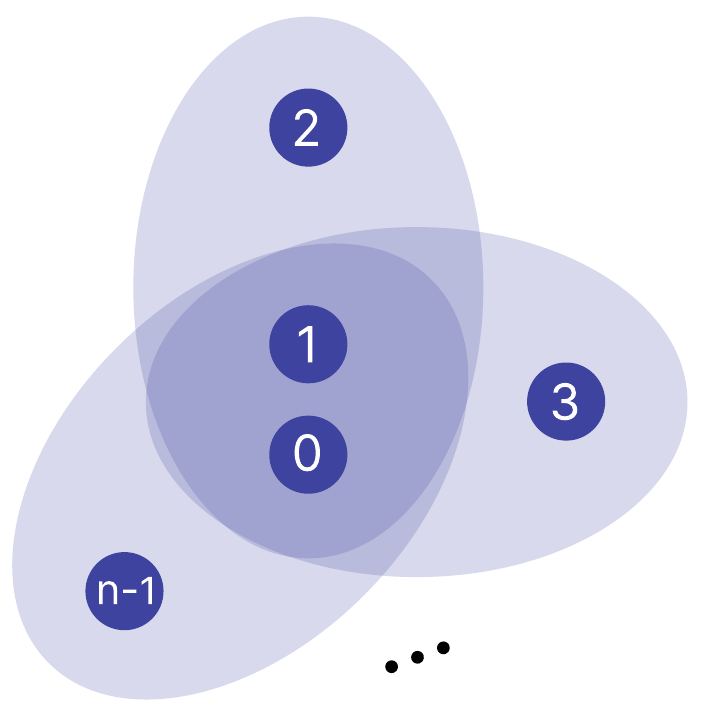}
    \caption{A 3-uniform star hypergraph with double-vertex overlap between adjacent hyperedges representing multiqubit states of structure (\ref{eq:star_double_state}).} \label{fig:star_double_hypergraphs}
\end{figure}

Similarly, quantum states encoding star hypergraphs are constructed. That is, we consider 3-uniform hypergraphs presented in Figures \ref{fig:star_single_hypergraphs}, \ref{fig:star_double_hypergraphs}, where all hyperedges are adjacent to each other and intersect over one or two vertices, forming star-like structures. Associated hypergraph states can be defined as follows 

\begin{align}
    |\psi_{HG}^{star, (1)}\rangle &=
    \prod_{k=0}^{m-1}\text{e}^{-\text{i} \theta_\mathbf{{k}} \sigma^y_{0} \sigma^z_{2k+1} \sigma^z_{2k+2}} |+\rangle^{\otimes n}, \label{eq:star_single_state}\\
    |\psi_{HG}^{star, (2)}\rangle &=
    \prod_{k=0}^{m-1}\text{e}^{-\text{i} \theta_\mathbf{{k}} \sigma^z_{0} \sigma^z_{1} \sigma^y_{k+2}} |+\rangle^{\otimes n}, \label{eq:star_double_state}
\end{align}
where $n$ is the number of vertices required to construct a star hypergraph of size $m$, which is equivalent to the number of qubits in the representative quantum system. In state (\ref{eq:star_single_state}) corresponding to stars with single-vertex overlap $n=2m+1$ and index $\mathbf{k} = (0, 2k+1, 2k+2)$, $k=0,...,m-1$, is used to iterate over the hyperedges. In state (\ref{eq:star_double_state}) representing stars with double-vertex overlap $n=m+2$, $\mathbf{k} = (0, 1, k+2)$, $k=0,...,m-1$. Note that to generate hypergraph states (\ref{eq:star_single_state}), we perform a series of conditional RY rotations over the state of the qubit representing the central vertex of the star in Figure \ref{fig:star_single_hypergraphs}. Qubits connected to the central one by corresponding hyperedges determine the directions of these rotations, and their magnitudes are given by hyperedge weights. On the contrary, for states (\ref{eq:star_double_state}) unitaries are selected in such a way that RY rotations are applied to the qubits at the extremities of the star, while qubits corresponding to the two central vertices in Figure \ref{fig:star_double_hypergraphs} control the directions of these rotations.

\begin{figure}[H]
    \centering
    \includegraphics[scale=0.3]{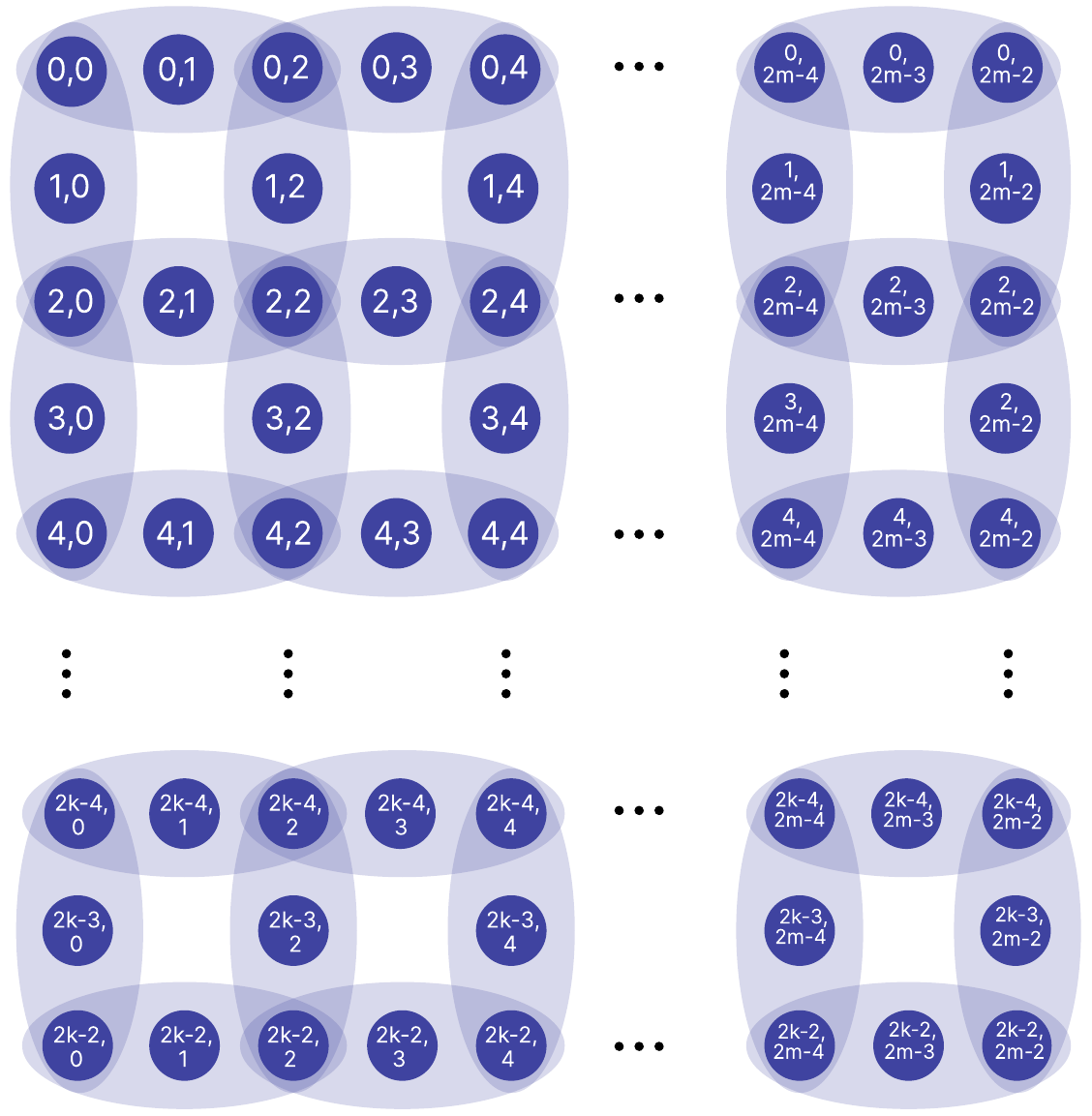}
    \caption{A 3-uniform hypergraph corresponding to a regular lattice with $k$ rows and $m$ columns, $k \geq 2$, $m \geq 2$, and representing multiqubit states of structure (\ref{eq:lattice_state_general}).}
    \label{fig:lattice_hypergraph}
\end{figure}

Further, we investigate hypergraph states of more complex structure, starting with regular lattices. The corresponding hypergraphs presented in Figure \ref{fig:lattice_hypergraph} are constructed in an ordered manner. Horizontal edges are first assembled from left to right to complete a lattice row, followed by a sequence of vertical edges attached to the row at regular intervals, forming segments of the lattice columns. These vertical edges connect the current row to the subsequent one, and the procedure is repeated iteratively until the lattice of the desired dimensions is complete. Note that a regular lattice hypergraph with $k$ rows and $m$ columns is composed of $k$ horizontal chain hypergraphs (see Figure \ref{fig:chain_single_hypergraphs}), each containing $2m-1$ vertices, and $m$ vertical chain hypergraphs, each containing $2k-1$ vertices, with the two sets of chains oriented orthogonally to each other. At the intersections, corresponding to lattice nodes, these chains form star-like structures (see Figure \ref{fig:star_single_hypergraphs}). Quantum states associated with such hypergraphs can be defined as follows

\begin{multline}
    |\psi_{HG}^{lattice}\rangle =
    \prod_{s=0}^{m-2} \left(\text{e}^{-\text{i} \theta_{2k-2, s} \sigma^z_{2k-2, 2s} \sigma^z_{2k-2, 2s+1} \sigma^y_{2k-2, 2s+2}}\right) \times \\ \times \prod_{r=0}^{k-2} \left(\prod_{q=0}^{m-1} \left(\text{e}^{-\text{i} \theta_{2r+1, q} \sigma^z_{2r, 2q} \sigma^z_{2r+1,2q} \sigma^y_{2r+2, 2q}}\right) \prod_{p=0}^{m-2} \left(\text{e}^{-\text{i} \theta_{2r, p} \sigma^z_{2r, 2p} \sigma^z_{2r,2p+1} \sigma^y_{2r, 2p+2}}\right) \right)|+\rangle^{\otimes n},
    \label{eq:lattice_state_general}
\end{multline}
where $k \geq 2$ and $m \geq 2$ are the numbers of rows and columns in the lattice, indices $r$ and $q$ run over the rows and columns, respectively, while indices $p$ and $s$ are used to iterate over the edges in individual rows. The number of vertices required to build a lattice of appropriate size, and hence the number of qubits in the representative quantum system, $n$ is equal to $(2m-1)k+(k-1)m$.

\begin{figure}[H]
    \centering
    \includegraphics[scale=0.25]{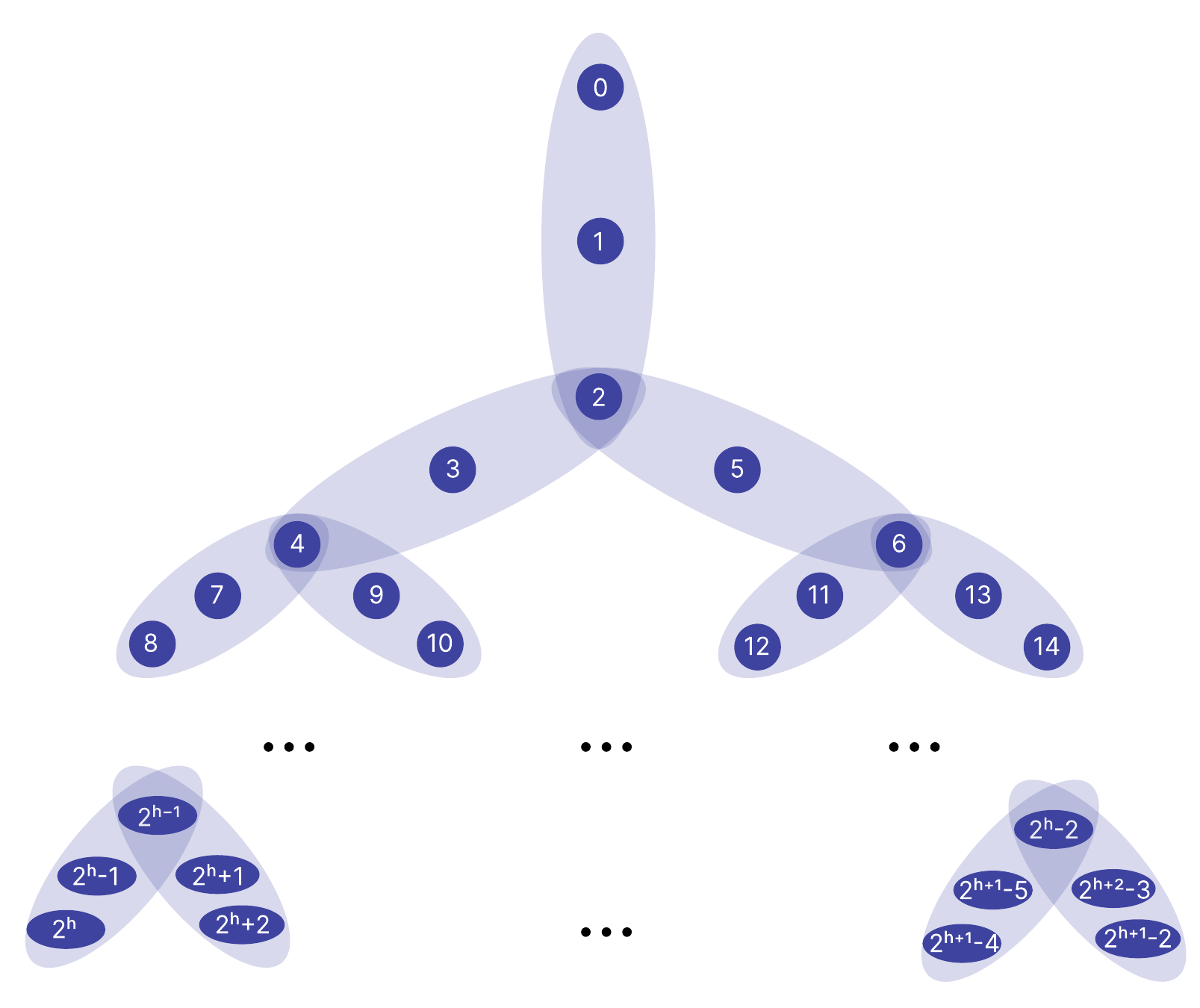}
    \caption{3-uniform hypergraph corresponding to a binary tree oh height $h$ and representing multiqubit states of structure (\ref{eq:tree_state_general}).}
    \label{fig:tree_hypergraph}
\end{figure}

Finally, hypergraph states associated with binary trees of arbitrary height are considered. Figure \ref{fig:tree_hypergraph} illustrates a representation of a binary tree as a 3-uniform hypergraph. The hypergraph is assembled level by level, resulting in a distinct ordering of hyperedges. The height of the tree $h$ is counted in hyperedges. Graph states constructed according to this connectivity pattern read

\begin{equation}
    |\psi_{HG}^{tree}\rangle =
    \prod_{r=1}^{h-1} \prod_{q=0}^{2^{r-1}-1} \prod_{p=0}^{1} \left(\text{e}^{-\text{i} \theta_\mathbf{{rqp}} \sigma^z_{2^r+2q} \sigma^z_{2^{r+1}+2(2q+p)- 1} \sigma^y_{2^{r+1} + 2(2q+p)}}\right)
    \text{e}^{-\text{i} \theta_{012} \sigma^z_{0} \sigma^z_{1} \sigma^y_{2}} |+\rangle^{\otimes n},
    \label{eq:tree_state_general}
\end{equation}
where $\mathbf{{rqp}} = (2^r+2q, 2^{r+1}+2(2q+p) - 1, 2^{r+1} + 2(2q +p))$, indices $r$, $q$, $p$ iterate over tree levels, couples of parent nodes within a specific level, and hyperedges adjacent to a specific couple of parent nodes, respectively; $n = 2^{h+1}-1$ is the number of vertices in the binary tree hypergraph of height $h \geq 2$. Corresponding number of hyperedges is equal to $2^{h}-1$.
Note that according to our convention, each hyperedge represents a connection between a couple of parent nodes (associated with Pauli-Z operators within the corresponding entangling unitary) and a child node (associated with Pauli-Y operator).

\section{Geometric Measure of Entanglement of Hypergraph States}\label{s2}
Let us investigate the entanglement of the introduced 3-uniform hypergraph states (\ref{eq:general_hypergraph_state}). We consider the geometric measure of entaglement defined as  (\ref{eq}).
 Importantly, the geometric measure of entanglement of a qubit with other qubits in a multipartite quantum state bears an explicit relation to the corresponding mean spin value, and can be expressed as follows

\begin{equation}
    E_l(|\psi\rangle) = \frac{1}{2}\left(1 - |\langle \boldsymbol{\sigma}_l\rangle|\right),
    \label{eq:geom_meas_ent_spin}
\end{equation}
where index $l$ denotes a qubit under consideration, $|\langle \boldsymbol{\sigma}_l\rangle| = |\langle \psi| \boldsymbol{\sigma}_l | \psi \rangle| = \sqrt{\langle \sigma^x_l \rangle^2 + \langle \sigma^y_l \rangle^2 + \langle \sigma^z_l\rangle^2 }$. In this interpretation, the geometric measure of entanglement quantifies bipartite entanglement between a single qubit and the remaining quantum system as a whole \cite{Samar}.

Relation (\ref{eq:geom_meas_ent_spin}) suggests that to quantify the entanglement of an arbitrary qubit $q_l$ with other qubits in state (\ref{eq:general_hypergraph_state}), it is enough to calculate expectation values of corresponding Pauli operators $\langle \sigma^x_l\rangle$, $\langle \sigma^y_l\rangle$, $\langle \sigma^z_l\rangle$. For 3-uniform hypergraph states generated by the action of RZZY operators and represented by chain (\ref{eq:chain_state_general}), star (\ref{eq:star_single_state}), (\ref{eq:star_double_state}), regular lattice (\ref{eq:lattice_state_general}) and binary tree (\ref{eq:tree_state_general}) hypergraphs, we obtain 

\begin{align}
        \langle \sigma_l^x\rangle &= \langle\psi_{HG}|\sigma^x_l|\psi_{HG}\rangle = \prod_{(i,j,k) \in E(l)} \cos{2\theta_{ijk}}, \label{eq:mean_spin_x}\\
        \langle \sigma_l^y\rangle &= \langle\psi_{HG}|\sigma^x_l|\psi_{HG}\rangle= 0,         \label{eq:mean_spin_y}\\
        \langle \sigma_l^z\rangle &= \langle\psi_{HG}|\sigma^x_l|\psi_{HG}\rangle= 0,
        \label{eq:mean_spin_z}
\end{align}
where $E(l) = \{(i,j,k)\in E | l \in (i,j,k)\}$ is a subset of hyperedges incident to the vertex with index $l$ representing qubit $q_l$, $E$ is a complete set of hyperedges, $\theta_{ijk}$ is the weight of hyperedge $(i,j,k)$. Substitution of (\ref{eq:mean_spin_x})-(\ref{eq:mean_spin_z}) into (\ref{eq:geom_meas_ent_spin}) yields

\begin{equation}
    E_l(|\psi_{HG}\rangle) = \frac{1}{2} \left(1 -\prod_{(i,j,k) \in E(l)} |\cos{2\theta_{ijk}}|\right).
    \label{eq:geom_meas_ent_hg}
\end{equation}
Based on this expression, we conclude that the geometric measure of entanglement of an arbitrary qubit with other qubits in weighted hypergraph states of specific structures (\ref{eq:chain_state_general})-(\ref{eq:tree_state_general}) is determined by the weights of all hyperedges containing the vertex representing this qubit in the hypergraph. Entanglement of qubit $q_l$ reaches its maximum value of $1/2$ in multiqubit states, where any of the state parameters $\theta_{ijk}$ corresponding to incident hyperedges $(i,j,k)\in E(l)$ is equal to $\pi/4$. The state of qubit $q_l$ is separable from the rest of the system if all the state parameters $\theta_{ijk}$, $(i,j,k)\in E(l)$, take values $\pi m/2$, $m \in \mathbb{Z}$.

Note that in the case where all the hyperedges are equally weighted, $\theta_{ijk}=\theta$, $(i,j,k) \in E$, the mean value of Pauli-X operator in the corresponding hypergraph state is reduced to

\begin{equation}
    \langle \sigma_l^x\rangle = \cos^{n_l}2\theta,
    \label{eq:mean_spin_x_equal_weights}
\end{equation}
resulting in

\begin{equation}
        E_l(|\psi_{HG}\rangle) = \frac{1}{2} \left(1-|\cos^{n_l}2\theta|\right),
    \label{eq:geom_meas_ent_hg_equal_weights}
\end{equation}
where $n_l$ is the degree of the vertex (number of its adjacent edges) associated with qubit $q_l$. Interestingly, expressions (\ref{eq:geom_meas_ent_hg}), (\ref{eq:geom_meas_ent_hg_equal_weights}) establish a connection between the entanglement, as a key physical property of the multiqubit states under investigation, and characteristics of hypergraphs, as mathematical objects used to represent these state.

\section{Quantifying the Entanglement of Hypergraph States with Quantum Computing}\label{s3}

The explicit relationship between the geometric measure of entanglement and mean spin (\ref{eq:geom_meas_ent_spin}) enables us to seamlessly estimate the entanglement of hypergraph states using quantum programming methods.

\begin{figure}[H]
    \centering
    \includegraphics[scale=0.8]{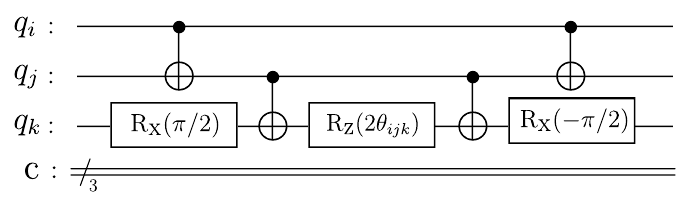}
    \caption{Gate-based representation of three-qubit unitary $\text{RZZY}_{ijk}(2\theta_{ijk})$ (\ref{eq:unitary}) composed of one-qubit rotation gates RX, RZ and two-qubit controlled gates CNOT.}
    \label{fig:protocol_rzzy}
\end{figure}

To begin with, let us develop a procedure for preparing arbitrary 3-uniform hypergraph states (\ref{eq:general_hypergraph_state}) on a universal gate-based quantum device. This procedure can be implemented in two stages, namely, initialization of the desired separable state of the multiqubit system and generation of interaction through the application of appropriate entangling unitaries. In our case, the n-qubit system corresponding to the quantum register of the device is prepared in a uniform superposition state using Hadamard gates. Note that traditionally, the quantum register is initialized in state $|0\rangle^{\otimes n}$ before the execution of any quantum program. Hence, we obtain

\begin{equation}
    |\psi_{init}\rangle = \prod_{l=0}^{n-1} H_l |0\rangle^{\otimes n} = |+\rangle^{\otimes n}.
\end{equation}
Subsequently, parameterized RZZY operators (\ref{eq:unitary}) are applied to each 3-qubit cluster associated with a subset of vertices incident to a common hyperedge of appropriate weight in the hypergraph representation of the target quantum state. Note that RZZY can be conveniently decomposed into a sequence of single-qubit parameterized rotation gates RX, RZ, and two-qubit controlled CNOT gates as shown in Figure \ref{fig:protocol_rzzy}. It is easy to verify that

\begin{equation}
    \text{RZZY}_{ijk}(2\theta_{ijk}) = \text{RX}_k(-\pi/2)\text{CNOT}_{ij}\text{CNOT}_{jk} \text{RZ}_k(2\theta_{ijk}) \text{CNOT}_{jk}\text{CNOT}_{ij}\text{RX}_k(\pi/2),
\end{equation}
which is consistent with identity (\ref{eq:rzzy_identity}). Here $\text{RX}_k(\pm \pi/2) = \exp(\mp\text{i} \pi \sigma^x_k/4) = \left(I_k \mp i \sigma^x_k\right)/\sqrt{2}$, $\text{RZ}_k(2\theta_{ijk}) = \exp(-\text{i}\theta_{ijk}\sigma^z_k)$ rotate the state of qubit $q_k$ around the X- and Z-axes by angles $\pm \pi/2$ and $2\theta_{ijk}$, respectively, $I_k$ is an identity gate acting on qubit $q_k$. Two-qubit gate $\text{CNOT}_{ij} = |0\rangle_i {}_{i}\langle0|\otimes I_j + |1\rangle_i {}_{i}\langle1|\otimes \sigma^x_j$ acts on states of qubits $q_i$, $q_j$ as a control and a target, respectively. The resulting multiqubit state reads

\begin{equation}
    |\psi_{HG}\rangle = \prod_{(i,j,k) \in E}RZZY_{ijk}(\theta_{ijk})\prod_{l=0}^{n-1} H_l |0\rangle^{\otimes n},
    \label{eq:hypergraph_state_qc}
\end{equation}
where the required number of qubits $n$ is determined by the number of vertices in the corresponding hypergraph, $E$ is a set of hyperedges, $\theta_{ijk}$ is the weight of hyperedge $(i,j,k)$.

To empirically quantify the entanglement of the generated hypergraph states employing quantum computations, we adhere to the well-established protocol. According to (\ref{eq:geom_meas_ent_spin}), the geometric measure of entanglement of qubit $q_l$ with the rest of the system is entirely determined by expectation values of observables $\sigma^x_l$, $\sigma^y_l$, $\sigma^z_l$. These expectation values are easily obtainable on a quantum computer based on the results of quantum measurements in the standard basis. Namely, we have

\begin{align}
    \langle \sigma^x_l \rangle &= \langle \psi_{HG} | \sigma^x_l | \psi_{HG} \rangle = \langle \tilde{\psi}^y_{HG}| \sigma^z_l | \tilde{\psi}^y_{HG} \rangle =|\langle \tilde{\psi}^y_{HG}| 0\rangle|^2 - |\langle \tilde{\psi}^y_{HG}| 1\rangle|^2, \label{eq:mean_x_meas}\\
    \langle \sigma^y_l \rangle &= \langle \psi_{HG} | \sigma^y_l | \psi_{HG} \rangle = \langle \tilde{\psi}^x_{HG}| \sigma^z_l | \tilde{\psi}^x_{HG} \rangle =|\langle \tilde{\psi}^x_{HG}| 0\rangle|^2 - |\langle \tilde{\psi}^x_{HG}| 1\rangle|^2, \label{eq:mean_y_meas}\\
    \langle \sigma^z_l \rangle &= \langle \psi_{HG} | \sigma^z_l | \psi_{HG} \rangle = |\langle \psi_{HG}| 0\rangle|^2 - |\langle \psi_{HG}| 1\rangle|^2, \label{eq:mean_z_meas}
\end{align}
where $|\tilde{\psi}^x_{HG} \rangle = \exp(-\textrm{i} \pi \sigma^x_l/4) |\psi_{HG}\rangle = RX_l(\pi/2) |\psi_{HG}\rangle$, $|\tilde{\psi}^y_{HG} \rangle = \exp(\textrm{i} \pi \sigma^y_l/4) |\psi_{HG}\rangle = RY_l(-\pi/2)|\psi_{HG}\rangle$. Expressions (\ref{eq:mean_x_meas}), (\ref{eq:mean_y_meas}) suggest that to facilitate the computation of $\langle \sigma^x_l \rangle$, $\langle \sigma^y_l \rangle$, the state of the target qubit $q_l$ should be rotated around the Y- and X-axes by angle $\pi/2$, respectively, before performing measurements in the standard basis.

\begin{figure}[H]
    \centering
    \includegraphics[scale=0.55]{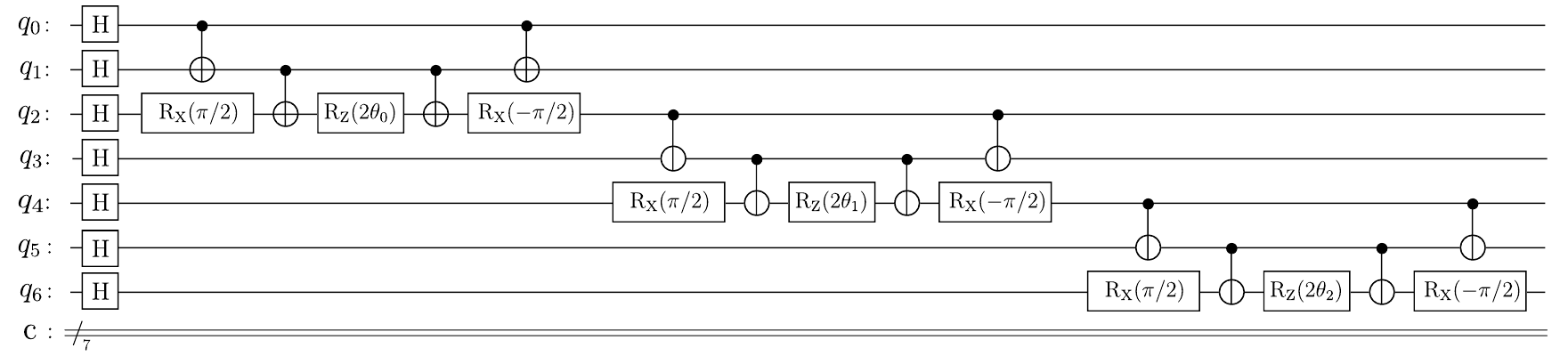}
    \caption{A protocol for preparing 7-qubit quantum hypergraph states (24) corresponding to the chain hypergraph with single-vertex intersection between consecutive hyperedges on a quantum universal gate-based device.
}
    \label{fig:protocol_chain}
\end{figure}

\begin{figure}[H]
  \centering
   \includegraphics[scale=0.5]{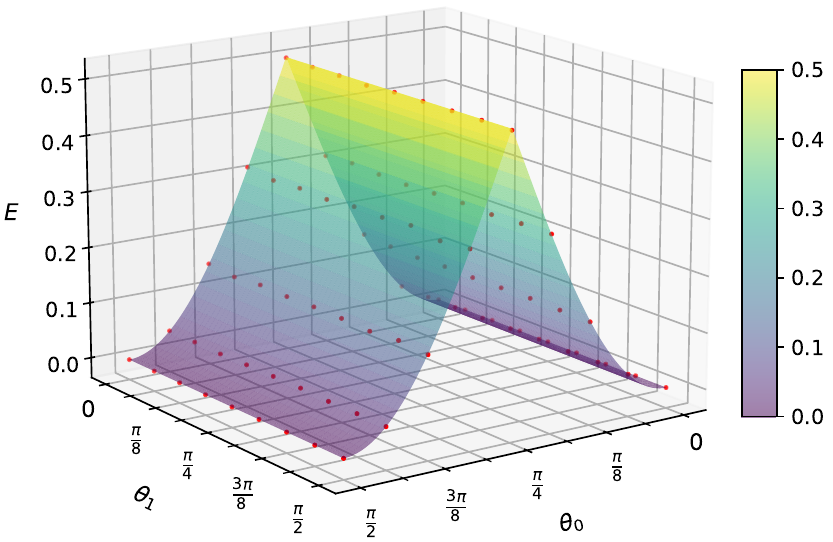}
    \caption{Geometric measure of entanglement of qubit $q_0$, corresponding to a hypergraph vertex of degree 1, with the remaining system in state (\ref{eq:chain_1_special}), where parameters $\theta_0$, $\theta_1$ vary independently, $\theta_2=\pi/8$.  Analytical dependencies on the state parameters are represented with a continuous surface. Results obtained with quantum computing on IBM’s AerSimulator for discrete parameter combinations are marked with dots.} \label{fig:results_chain_single_deg_1}
\end{figure}

\begin{figure}[H]
  \centering
   \includegraphics[scale=0.5]{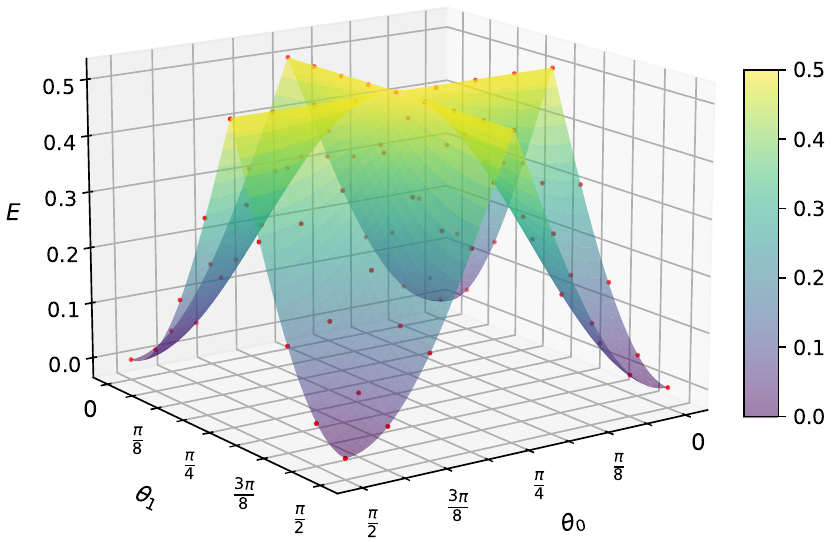}
    \caption{Geometric measure of entanglement of qubit $q_1$, corresponding to a hypergraph vertex of degree 2, with the remaining system in state (\ref{eq:chain_1_special}), where parameters $\theta_0$, $\theta_1$ vary independently, $\theta_2=\pi/8$.  Analytical dependencies on the state parameters are represented with a continuous surface. Results obtained with quantum computing on IBM’s AerSimulator for discrete parameter combinations are marked with dots.} \label{fig:results_chain_single_deg_2}
\end{figure}

\begin{figure}[H]
    \centering    
    \includegraphics[scale=0.5]{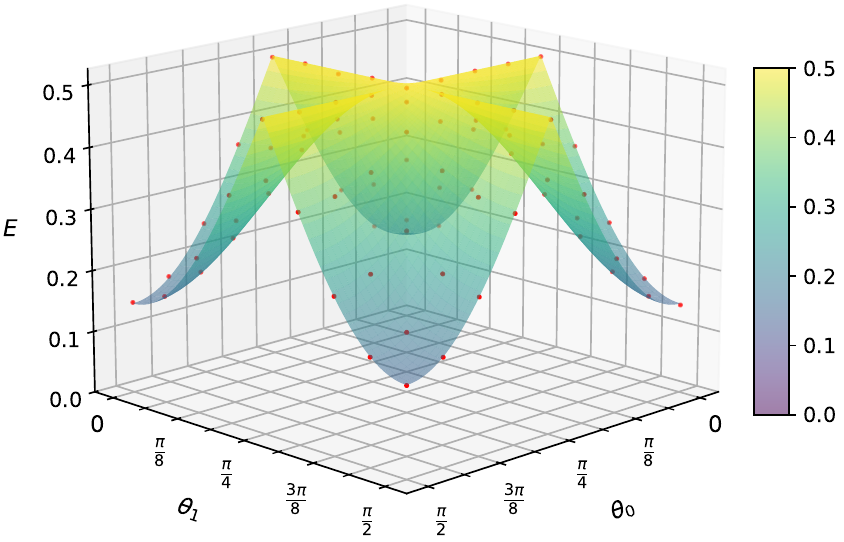}
    \caption{Geometric measure of entanglement of qubit $q_2$, corresponding to a hypergraph vertex of degree 3, with the remaining system in state (\ref{eq:chain_2_special}), where parameters $\theta_0$, $\theta_1$ vary independently, $\theta_2=\pi/8$.  Analytical dependency on the state parameters is represented with a continuous surface. Results obtained with quantum computing on  IBM’s AerSimulator for discrete parameter combinations are marked with dots.
}
    \label{fig:results_chain_double}
\end{figure}

We next apply the outlined methodology to prepare and examine special cases of 3-uniform edge-ordered hypergraph states represented by complex structures described in Section 2. Corresponding quantum protocols for multiqubit state generation and entanglement quantification are implemented using IBM's high-performance quantum simulator AerSimulator in an ideal setting with no specified noise model. Namely, we consider subclasses of states (\ref{eq:general_hypergraph_state}) associated with chain (\ref{eq:chain_state_general}), star (\ref{eq:star_single_state}), (\ref{eq:star_double_state}), regular lattice (\ref{eq:lattice_state_general}) and binary tree (\ref{eq:tree_state_general}) hypergraphs. In the first two cases, hypergraphs with different connectivity patterns are examined, where adjacent hyperedges of size 3 intersect over one or two vertices. Firstly, we construct quantum states (\ref{eq:chain_state_general}), $\alpha=1,2$, corresponding to chain hypergraphs of size 3 

\begin{align}
    |\psi_{HG}^{chain,(1)}\rangle &= 
    \prod_{k=0}^{2}\text{e}^{-\text{i} \theta_\mathbf{{k}} \sigma^z_{2k} \sigma^z_{2k+1} \sigma^y_{2k+2}} |+\rangle^{\otimes 7}, \label{eq:chain_1_special}\\
    |\psi_{HG}^{chain,(2)}\rangle &= 
    \prod_{k=0}^{2}\text{e}^{-\text{i} \theta_\mathbf{{k}} \sigma^z_{k} \sigma^z_{k+1} \sigma^y_{k+2}} |+\rangle^{\otimes 5}, \label{eq:chain_2_special}
\end{align}
where the required number of qubits in the system depends on the type of connectivity in the chain. As an example, Figure \ref{fig:protocol_chain} shows a protocol for preparing state (\ref{eq:chain_1_special}) on a quantum device. The remaining states considered in this study are prepared similarly, taking into account the structure of corresponding hypergraphs. For instance, we also examine states
\begin{align}
    |\psi_{HG}^{star, (1)}\rangle &=
    \prod_{k=0}^{2}\text{e}^{-\text{i} \theta_\mathbf{{k}} \sigma^y_{0} \sigma^z_{2k+1} \sigma^z_{2k+2}} |+\rangle^{\otimes 7}, \label{eq:star_1_special}\\
    |\psi_{HG}^{star, (2)}\rangle &=
    \prod_{k=0}^{2}\text{e}^{-\text{i} \theta_\mathbf{{k}} \sigma^z_{0} \sigma^z_{1} \sigma^y_{k+2}} |+\rangle^{\otimes 5},\label{eq:star_2_special}
\end{align}
belonging to subclasses (\ref{eq:star_single_state}), (\ref{eq:star_double_state}) represented by star hypergraphs of the same size with one and two central vertices, where all three hyperedges intersect. We further detect the geometric measure of entanglement of qubits corresponding to hypergraph vertices of different degrees with the remaining system in states (\ref{eq:chain_1_special})-(\ref{eq:star_2_special}) and study its dependency on state parameters introduced through the action of entangling RZZY unitaries and representing the weights of associated hyperedges. To this end, parameters $\theta_0$, $\theta_1$ are varied independently in the range $[0, \pi/2]$, covering one period of function (\ref{eq:geom_meas_ent_hg}), while parameter $\theta_2$ is fixed at $\pi/8$. Corresponding results obtained from quantum simulations for the chain hypergraph states are presented in Figures \ref{fig:results_chain_single_deg_1}-\ref{fig:results_chain_double}. Specifically, we quantify the entanglement of qubits $q_0$, $q_1$ represented by vertices of degree 1 and 2 (see Figures \ref{fig:results_chain_single_deg_1}, \ref{fig:results_chain_single_deg_2}, respectively) in hypergraph state (\ref{eq:chain_1_special}) corresponding to the chain with single-vertex overlap, as well as qubit $q_2$ represented by vertex of degree 3 in state (\ref{eq:chain_2_special}) corresponding to the chain with double-vertex overlap (see Figure \ref{fig:results_chain_double}). It is easy to see that the geometric measure of entanglement of each qubit depends solely on the weights of incident hyperedges. Note that, since parameter $\theta_2$ in state (\ref{eq:chain_2_special}) takes constant value of $\pi/8$, the geometric measure of entanglement always exceeds zero for qubit $q_2$, and thus, it is always entangled with the rest of the system. As expected from relationship (\ref{eq:geom_meas_ent_hg}), similar results were obtained for states (\ref{eq:star_1_special})-(\ref{eq:star_2_special}) represented by star structures, where qubits are associated with vertices of degree 1 (vertices at the star extremities) or 3 (central vertices).

\begin{figure}[H]
    \centering    
    \includegraphics[scale=0.5]{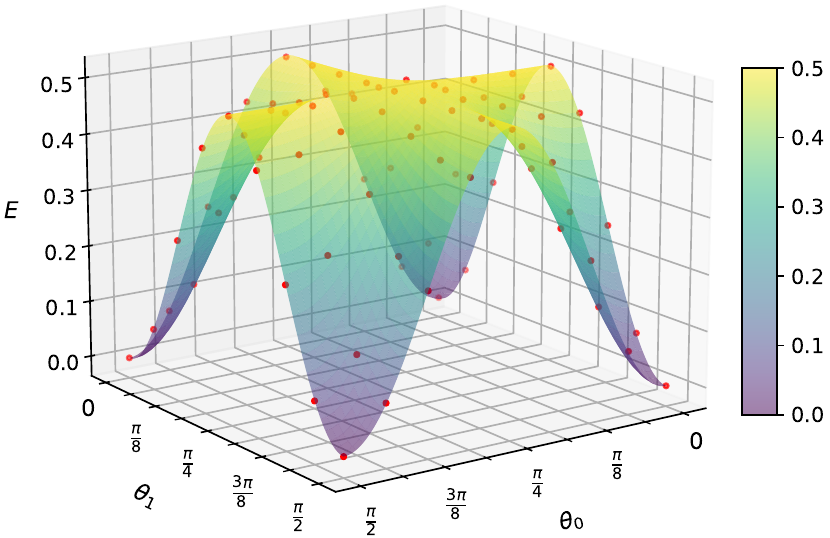}
    \caption{Geometric measure of entanglement of qubit $q_{2,2}$, corresponding to a hypergraph vertex of degree 4, with the remaining system in state (\ref{eq:lattice_state_special}), where parameters $\theta_0$, $\theta_1$ vary independently.  Analytical dependency on the state parameters is represented with a continuous surface. Results obtained with quantum computing on  IBM’s AerSimulator for discrete parameter combinations are marked with dots.}
    \label{fig:results_lattice}
\end{figure}

Additionally, we consider a special case of state (\ref{eq:lattice_state_general}) represented by a regular lattice with 3 rows and 3 columns 

\begin{multline}
    |\psi_{HG}^{lattice}\rangle =
    \prod_{s=0}^{1} \left(\text{e}^{-\text{i} \theta_{0} \sigma^z_{4, 2s} \sigma^z_{4, 2s+1} \sigma^y_{4, 2s+2}}\right) \times \\ \times \prod_{r=0}^{1} \left(\prod_{q=0}^{2} \left(\text{e}^{-\text{i} \theta_{1} \sigma^z_{2r, 2q} \sigma^z_{2r+1,2q} \sigma^y_{2r+2, 2q}}\right) \prod_{p=0}^{1} \left(\text{e}^{-\text{i} \theta_{0} \sigma^z_{2r, 2p} \sigma^z_{2r,2p+1} \sigma^y_{2r, 2p+2}}\right) \right)|+\rangle^{\otimes 21},
\label{eq:lattice_state_special}
\end{multline}
where all hyperedges forming the rows are assigned the weight $\theta_0$, while all the hyperedges forming columns are assigned the weight $\theta_1$. Essentially, this means that the strength of 3-partite interaction in the qubit chains depends on the direction within the lattice. In this state, we study the geometric measure of entanglement of qubit $q_{2,2}$ corresponding to the central vertex of the lattice (degree 4) with other qubits. The results of this investigation are presented in Figure \ref{fig:results_lattice}.

\begin{figure}[H]
    \centering    
    \includegraphics[scale=0.3]{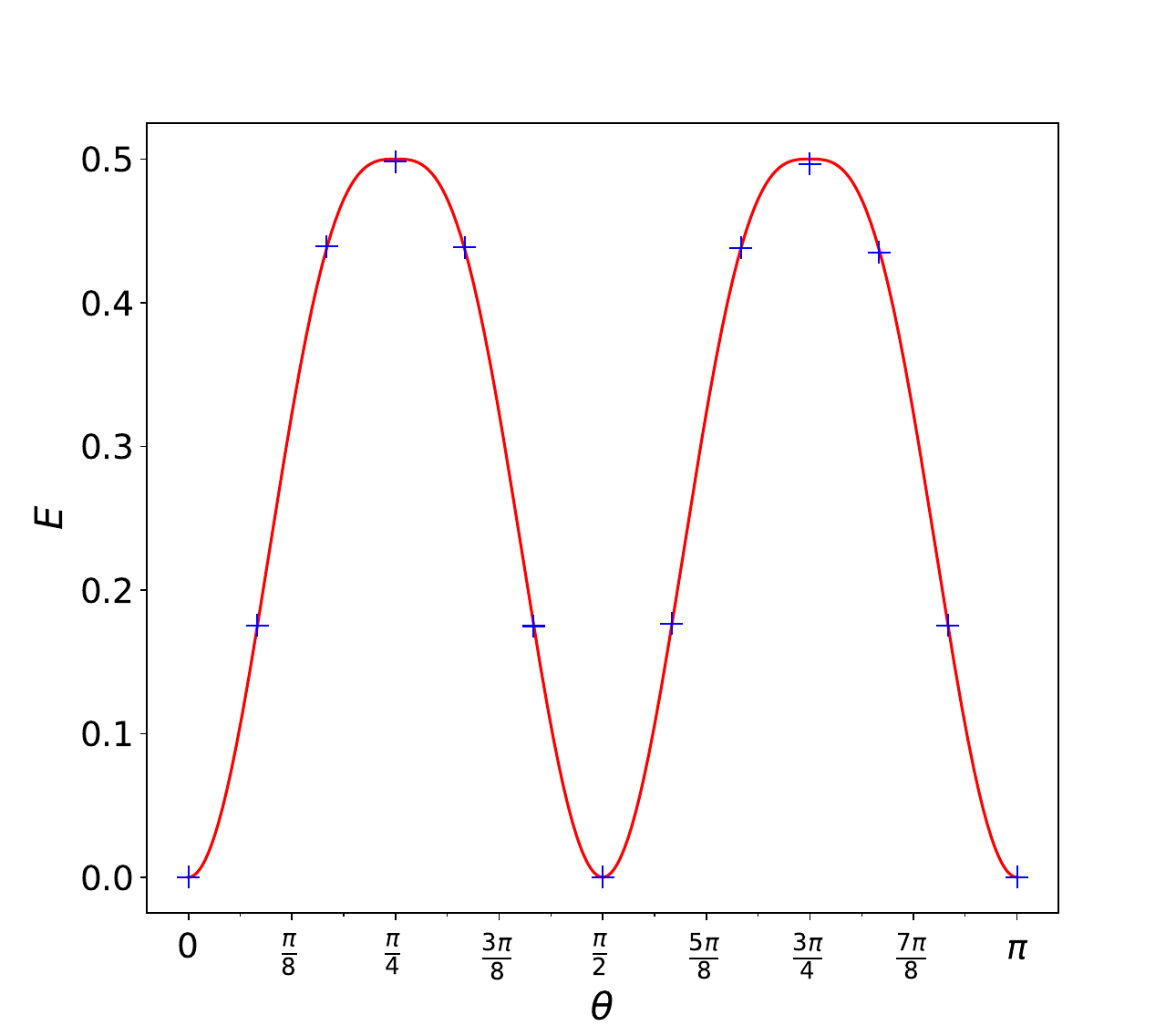}
    \caption{Geometric measure of entanglement of qubit $q_4$, corresponding to a hypergraph vertex of degree 3, with the remaining system in state (\ref{eq:tree_state_special}).  Analytical dependency on the state parameter $\theta$ is represented with a continuous curve. Results obtained with quantum computing on IBM’s AerSimulator for discrete parameter values are marked with crosses.}
    \label{fig:results_tree}
\end{figure}

Finally, a 3-uniform hypergraph state corresponding to the binary tree of height 4 is examined. Namely, we have

\begin{equation}
    |\psi_{HG}^{tree}\rangle =
    \prod_{r=1}^{3} \prod_{q=0}^{2^{r-1}-1} \prod_{p=0}^{1} \text{e}^{-\text{i} \theta \sigma^z_{2^r+2q} \sigma^z_{2^{r+1}+2(2q+p)- 1} \sigma^y_{2^{r+1} + 2(2q+p)}}
    \text{e}^{-\text{i} \theta \sigma^z_{0} \sigma^z_{1} \sigma^y_{2}} |+\rangle^{\otimes 31},
    \label{eq:tree_state_special}
\end{equation}
where all the weights of the hypergraph are set equal to each other and denoted by $\theta$. The results of quantifying the geometric measure of entanglement of qubit $q_4$ corresponding to the tree node of degree 3 are presented in Figure \ref{fig:results_tree}.

As evident from Figures \ref{fig:results_chain_single_deg_1}-\ref{fig:results_tree}, for all the hypergraph states investigated, the entanglement estimates obtained on the basis of direct mean spin measurements are consistent with analytical predictions given by expression (\ref{eq:geom_meas_ent_hg}). This result demonstrates the potential of quantum computing methods for studying properties of multipartite quantum states corresponding to complex structures.

\section{Conclusions}\label{s4}
A method for constructing 3-uniform edge-ordered hypergraph quantum states has been introduced. The states are prepared using parameterized three-qubit $RZZY$ entangling operators. The noncommutativity of these operators establishes the correspondence between the constructed quantum states and edge-ordered hypergraphs. The hyperedge weights are encoded in the parameters of the operators. An analytical expression for the geometric measure of entanglement of a qubit with the remaining qubits in 3-uniform edge-ordered hypergraph quantum states has been derived. It has been shown that for a selected qubit the entanglement is determined by the parameters of the hyperedges incident to the corresponding vertex of the hypergraph. Thus, a direct connection between a property of the quantum state, namely its entanglement, and the local structural characteristics of the corresponding hypergraph has been established. In the special case of equally weighted hyperedges, the entanglement has been shown to depend on the vertex degree (\ref{eq:geom_meas_ent_hg_equal_weights}).

To demonstrate the relationship between local connectivity in hypergraphs and the entanglement of the corresponding qubits in hypergraph quantum states, several 3-uniform hypergraph structures were considered, including chains with single- and double-vertex intersections, star hypergraphs, a regular lattice, and a binary tree. Quantum protocols for preparing the considered hypergraph states and quantifying their entanglement were constructed. The entanglement was quantified using IBM's AerSimulator (see Figs. \ref{fig:results_chain_single_deg_1}--\ref{fig:results_tree}). Corresponding entanglement values obtained from the quantum simulations agree with the analytical dependencies.

The results demonstrate that the entanglement of the considered multiqubit states is directly related to the local hypergraph structure and hyperedge weights, providing a framework for studying hypergraph properties using quantum computing. Also, the analytical dependencies obtained for the entanglement of multiqubit hypergraph states in this paper can serve as a basis for preparing quantum states with specified entanglement properties. Additionally, they can be used in further studies of relevant quantum information problems.

  \end{document}